\documentclass[aps,preprint,superscriptaddress,groupedaddress]{revtex4}  
\usepackage{graphicx}  
\usepackage{dcolumn}   
\usepackage{bm}        
\usepackage{amssymb}   
\usepackage{amsmath}
\usepackage{bbm}
\usepackage{doi}
\usepackage{hyperref}
\usepackage{epsfig}
\usepackage[section]{placeins}
\usepackage{float}
\usepackage{stackengine}
\setstackEOL{\#}
\setstackgap{L}{12pt}

\usepackage{color}
\usepackage{soul}

\begin{document}
\title{Bistable emergence of oscillations in structured cell populations} 

\author{Rosa Martinez-Corral}
\affiliation{Department of Experimental and Health Sciences, Universitat Pompeu Fabra, Barcelona, Spain}
\author{Jintao Liu}
\affiliation{Center for Infectious Disease Research and Tsinghua-Peking Center for Life Sciences, School of Medicine, Tsinghua University, Beijing, China}
\author{Gurol Suel}
\affiliation{Division of Biological Sciences, University of California San Diego, California 92093, USA}
\affiliation{San Diego Center for Systems Biology, University of California San Diego, California 92093, USA}
\affiliation{Center for Microbiome Innovation, University of California San Diego, California 92093, USA}
\author{Jordi Garcia-Ojalvo}
\affiliation{Department of Experimental and Health Sciences, Universitat Pompeu Fabra, Barcelona, Spain}
\date{\today}

\keywords{Biological oscillations, delay-induced oscillations, subcritical Hopf bifurcation, delayed negative feedback, biofilm growth}

\begin{abstract}

Biofilm communities of \emph{Bacillus subtilis} bacteria have recently been shown to exhibit collective growth-rate oscillations mediated by electrochemical signaling to cope with nutrient starvation. These oscillations emerge once the colony reaches a large enough number of cells. However, it remains unclear whether the amplitude of the oscillations, and thus their effectiveness, builds up over time gradually, or if they can emerge instantly with a non-zero amplitude. Here we address this question by combining microfluidics-based time-lapse microscopy experiments with a minimal theoretical description of the system in the form of a delay-differential equation model. Analytical and numerical methods reveal that oscillations arise through a subcritical Hopf bifurcation, which enables instant high amplitude oscillations. Consequently, the model predicts a bistable regime where an oscillating and a non-oscillating attractor coexist in phase space. We experimentally validate this prediction by showing that oscillations can be triggered by perturbing the media conditions, provided the biofilm size lies within an appropriate range. The model also predicts that the minimum size at which oscillations start decreases with stress, a fact that we also verify experimentally. Taken together, our results show that collective oscillations in cell populations can emerge suddenly with non-zero amplitude via a discontinuous transition.

\end{abstract}
\maketitle






\section*{Introduction}
One of the main defining features of collective self-organization in coupled dynamical systems is the requirement of a minimum system size \cite{Winfree:2002tg}.
According to that scenario, the number of coupled elements behaves as a control parameter that has to exceed a certain threshold for collective behavior to arise \cite{Zamora-Munt:2010cr}.
Evidence of biological processes requiring a critical cell density has been reported for instance in myoblast fusion \cite{Konigsberg:1971mz}, yeast glycolysis \cite{Aldridge:1976uq}, amoebae aggregation initiation \cite{Gregor:2010qf}, and immune cell homeostasis \cite{Hart:2014nx}, among others.
The phenomenon also underlies the most commonly studied means of cell-cell communication in bacteria, namely quorum sensing \cite{Garcia-Ojalvo:2004kl,Ng:2009gf,Danino:2010oq}.
In fact, the emergence of synchronized oscillations due to coupling in general systems of interacting elements has been termed {\em dynamical quorum sensing} \cite{De-Monte:2007fk,Taylor:2009dq}, due to its conceptual links with bacterial communication.

\begin{figure}[htb]
\centerline{\includegraphics[width=0.65\textwidth]{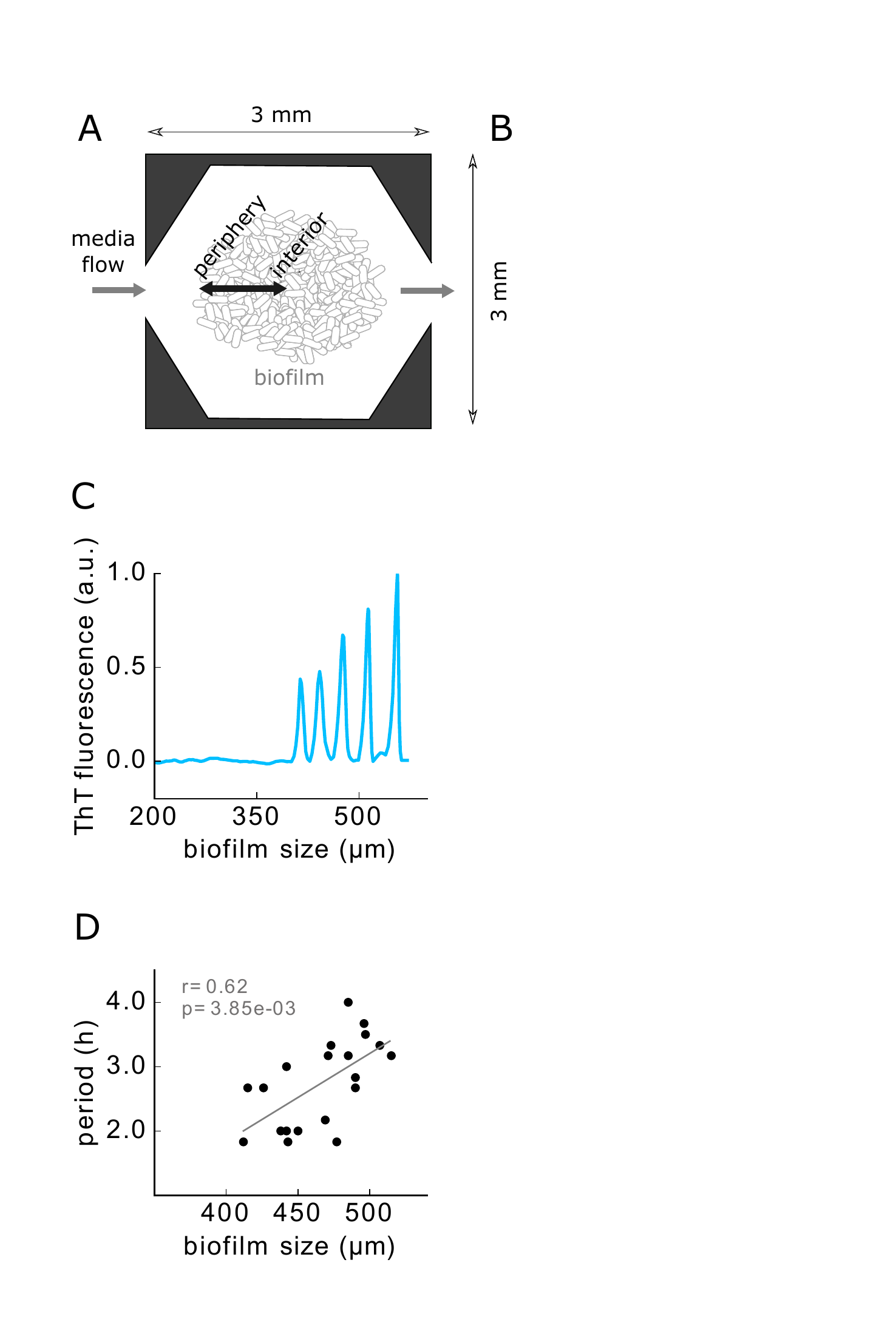}}
 \caption{Emerging oscillations in \emph{Bacillus subtilis} biofilms.
 (A) Scheme of the microfluidic device used to grow biofilms.
 (B) Filmstrip showing oscillations in stress reported by the membrane potential marker Thioflavine T (ThT).
 (C) ThT signal as the biofilm grows in size (radius), showing that the oscillations start only after a critical size is reached.
 (D) Period at the oscillation onset as a function of biofilm size (radius). Scatterplot: experimental data; grey line: linear fit; r: correlation coefficient; p: p-value.}
 \label{fig1}
\end{figure}

In the studies of dynamical quorum sensing carried out so far, coupling usually connects all elements of the system in an all-to-all manner (global coupling), so that the interaction between one cell and the rest can be described by a mean-field approximation \cite{Gregor:2010qf}.
Such a global-coupling approximation may not be valid, however, when the population is structured in space. This is the case, in particular, when communication between cells is not mediated by a rapidly diffusive signal, but by a pulse-coupling mechanism through which the cells become activated by their neighbours in a ``bucket-brigade'' manner.
This happens for instance in neurons, which are coupled via chemical synapses, and, as we have reported recently, in bacterial biofilms \cite{Liu:2015fk}, where the cells signal their immediate neighbors via potassium ions released by bacterial ion channels \cite{Prindle:2015uq}.
In this latter case, electrical signaling enables bacteria in the interior of the biofilm (which are subject to severe limitation in their only nitrogen source, glutamate) to transmit their stress state to the cells in the periphery (Fig.~\ref{fig1}A).
Peripheral cells subsequently stop growing and allow glutamate to enter the biofilm, thereby releasing the stress in the interior \cite{Liu:2015fk}.
This constitutes a spatially distributed negative feedback, which acts with a non-negligible delay due to the relatively slow propagation of the stressor.

Delayed negative feedback is well known to induce oscillations (or even more complex aperiodic dynamical phenomena), in particular in biological systems such as gene regulatory networks \cite{Smolen:2003kx,Bratsun:2005fk,Novak:2008vn,Mather:2009uq}, physiological control systems \cite{Batzel:2011zr,Mackey:1977ys}, neuronal populations \cite{Lindner:2005ve}, and ecological communities \cite{May:1973qf}.
In \emph{Bacillus subtilis} biofilms, the above-described delayed negative feedback leads to oscillations in growth and stress levels \cite{Liu:2015fk}.
This is shown in Fig.~\ref{fig1}B, which depicts a biofilm growing under constant conditions in a microfluidic system (Fig.~\ref{fig1}A) \cite{Liu:2015fk}.
In the filmstrip, the blue color represents light emitted by the fluorescent cationic dye Thioflavin T (ThT), which acts as a Nernstian voltage indicator that reports on the cellular membrane potential (cells light up when hyperpolarized) \cite{Prindle:2015uq,Humphries:2017kx}.
Hyperpolarization results from intracellular potassium release, which is in turn caused by stress \cite{Prindle:2015uq}.
Thus ThT is a reporter of cellular stress, and the ThT oscillations portrayed in Fig.~\ref{fig1}B can therefore be considered periodic modulations in stress.

The oscillations described above exhibit two distinct characteristics.
First, they appear only after the biofilm has reached a minimum size \cite{Liu:2015fk}.
This can be observed in Fig.~\ref{fig1}C, which shows the ThT signal as a function of biofilm size  (corresponding roughly to a time series, since system size increases monotonically with time).
Second, the period of the oscillations increases with the size of the biofilm.
This can be seen in Fig.~\ref{fig1}D, which plots the period at oscillation onset, as a function of the size of the biofilm at that time.
Here we propose a basic explanation of these two facts in terms of a minimal delay-differential equation (DDE) model, in which the delay is considered explicitly.
We show that a single DDE representing the stress dynamics is able to explain both the experimentally observed emergence of biofilm oscillations at a critical size and the dependence of their period on the biofilm size.
The model makes two testable predictions.
First, oscillations emerge via a subcritical Hopf bifurcation, which implies the existence of a bistable region where a stable state of steady growth coexists with the oscillatory regime described above.
Second, the critical system size at which oscillations start decreases with nutritional stress.
We validate experimentally these two expectations, thereby supporting the hypothesis that stress oscillations in growing biofilms emerge through a delay-induced Hopf instability.

\section*{A DDE model for biofilm stress dynamics}

\begin{figure}[htb]
\centerline{\includegraphics[width=0.45\textwidth]{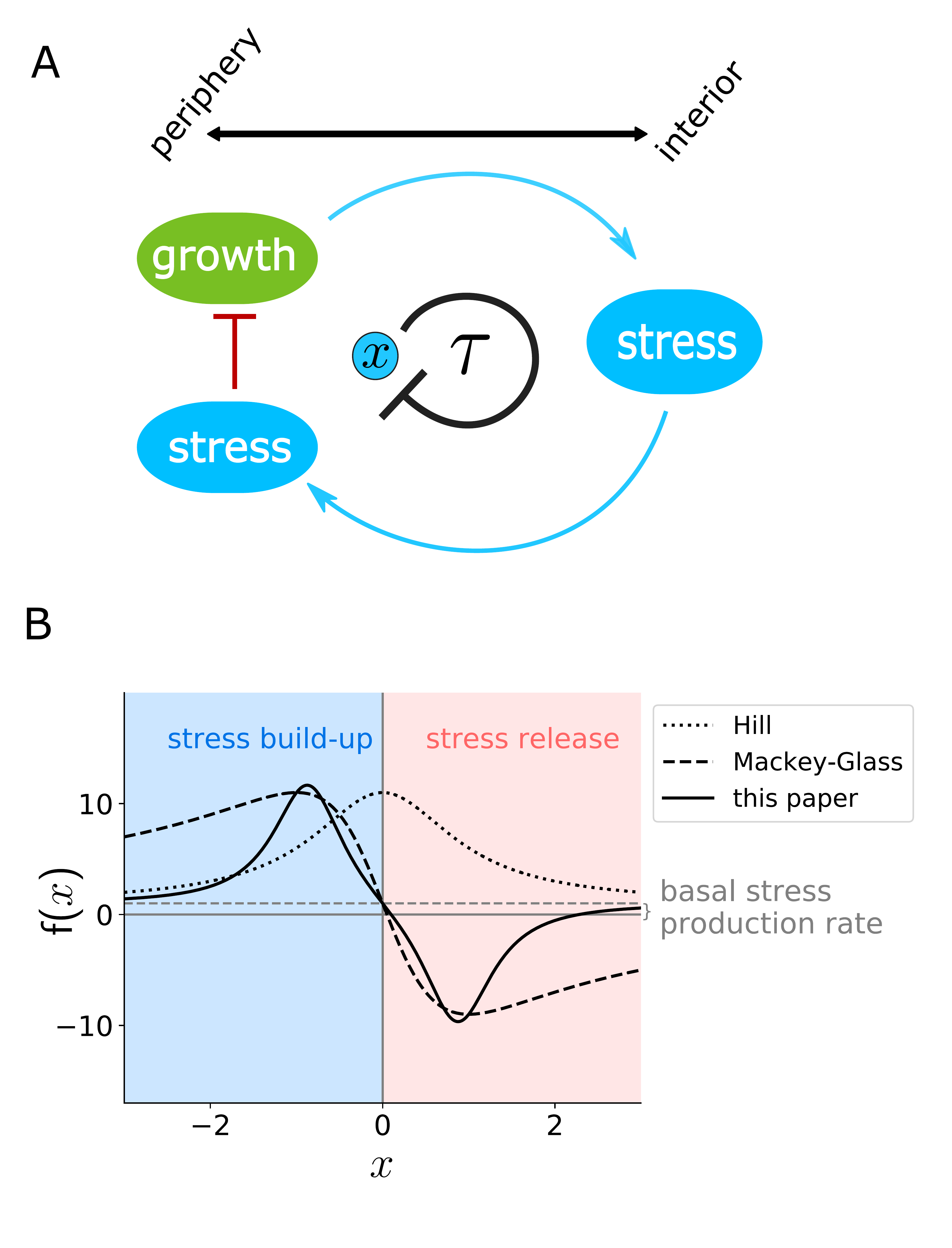}}
 \caption{A DDE model of biofilm oscillations. (A) Scheme of the model. Growth in the periphery leads to metabolic stress in the interior, which is transmitted to the periphery and inhibits growth. Therefore, stress in the periphery ($x$) is considered to exert a negative feedback on itself with some delay ($\tau$).
 (B) Stress production function $f(x)$ for $C=1, \alpha=10, \beta=\gamma=1$. The dotted grey line marks $f(x)=C$. Dotted  black line: Hill function $1+\frac{10}{1+x^2}$. Dashed line: Mackey-Glass function $1-\frac{20x}{1+x^2}$.
 }\label{fig2}
\end{figure}

We set to build a dynamical model describing the behavior of the stress levels in the biofilm periphery, which we denote by $x(t)$.
This variable is taken to represent the stress of the peripheral cells with respect to a baseline, such that it can take positive or negative values depending on whether stress levels are above or below this baseline.
In its simplest form, we can assume that the dynamics of $x(t)$ is given by a production rate that depends on past stress levels, $f(x(t-\tau))$, and a linear degradation with rate $\delta$:
\begin{align}
\frac{dx}{dt}=f(x(t-\tau))-\delta x\,.
\label{eq:dde}
\end{align}
To choose $f(x)$, we assume that peripheral cells always have some basal stress production $C$ that is modified by the effect of the stress experienced by the periphery in the past (Fig.~\ref{fig2}A).
Up to a certain level, peripheral stress at a time $\tau$ ago, $x(t-\tau)$, would have halted growth and thus it would now, at time $t$, lead to a reduction in stress levels coming from the interior ($f(x)<C$).
On the other hand, if stress was too high, the interior cells would have probably died, thus eliminating the feedback effect ($f(x)=C$).
Conversely, past peripheral stress levels below baseline would have allowed growth at time $t-\tau$, leading to an increase in stress at time $t$ ($f(x)>C$).
Again, values of $x$ well below the baseline would be indicative of no stress at all, and thus no feedback would be present ($f(x)=C$). A simple functional form fulfilling all these requirements is (solid line in Fig.~\ref{fig2}B):
\begin{align}
f(x)=C-\frac{\alpha x}{1-(x/\beta)^{2} + (x/\gamma)^{4}}
\label{eq:fx}
\end{align}
Note that a simple negative feedback expression like the Hill function ${1}/{(1+x^2)}$, which is often used to model self-inhibition \cite{Lema:2000xy,Suzuki:2016fk}, is not appropriate here since it cannot change sign, and thus it would only allow for either stress release or build-up, irrespective of the amount of stress the biofilm had in the past.
Another common representation of feedback is given by the Mackey-Glass function ${x}/(1+x^2)$ \cite{Mackey:1977ys}.
We do not use that functional form here either, since $f(x)=C-{x}/(1+x^2)$ would decay too slowly (dashed line in Fig.~\ref{fig2}B) for very high and very low stress; we expect biofilm viability and metabolic feedback to be lost much more quickly if stress becomes too high or too low, respectively.
To reproduce such quickly decay to the baseline, we use the functional form given in Eq.~\ref{eq:fx}, in which a fourth-order term is included in the denominator.
We have also chosen a negative sign in front of the second-order term ($\beta x^2$).
This ensures that $f(x)$ depends more weakly on the stress for smaller stress levels than for stronger ones (compare the solid and dashed lines in the negative-slope region of Fig.~\ref{fig2}B).

The time-delay parameter in equation \ref{eq:dde} includes the time needed for the growth state of the periphery to affect the stress of the interior cells, and the time that it takes for the stress signal coming from the interior to reach the peripheral cells (Fig.~\ref{fig2}A).
We now examine how this delay term affects the dynamical behavior of the system.

\section*{Delayed negative feedback leads to oscillations via a Hopf bifurcation}
\label{stabilityanalysis}

Equation~\ref{eq:dde} has a steady-state solution, given implicitly by:
\begin{align}
C -  \frac{\alpha\,x_{s}}{1-x_{s}^{2} + x_{s}^{4}}-\delta x_{s} =0\label{eq:stdst}
\end{align}
(we assume in what follows that $\beta=\gamma=1$).
The stability of this stationary state is determined by introducing the ansatz $x(t)=x_s+\delta x\exp(\lambda t)$ into Eq.~\ref{eq:dde} and expanding $f(x)$ up to first order in $\delta x$. This leads to the characteristic equation \cite{erneux2009applied}
\begin{align}
J_\tau\exp(-\lambda\,\tau)+J_0-\lambda=0\,,
\label{chareq}
\end{align}
where $\lambda\equiv\mu+i\nu$ is the complex eigenvalue of the steady-state solution, $J_0=-\delta$ is the derivative of the right-hand side of Eq.~\ref{eq:dde} with respect to the non-delayed variable $x(t)$, and $J_{\tau}$ is the derivative with respect to the delayed variable $x_\tau\equiv x(t-\tau)$, evaluated at the steady state:
\begin{align}
J_{\tau}=- \frac{\alpha\,x_{s} \left(2 x_s-4 x_{s}^{3}\right)}{\left(1-x_{s}^{2} + x_{s}^{4}\right)^{2}} - \frac{\alpha}{1-x_{s}^{2} + x_{s}^{4}}
\label{eq:Jtau}
\end{align}
Expanded into its real and imaginary parts, the characteristic equation \ref{chareq} takes the form:
\begin{align}
&J_\tau\exp(-\mu\tau)\cos(\nu\tau)+J_0-\mu =0 \label{eq1:real} \\
&J_\tau\exp(-\mu\tau)\sin(\nu\tau)+\nu=0 \label{eq2:im}
\end{align}
A bifurcation in which the steady state given by Eq.~\ref{eq:stdst} changes stability entails that $\mu$ goes through 0.
Setting $\mu$ to 0 in Eqs.~\ref{eq1:real} and \ref{eq2:im} leads to the following solution for $\nu$:
\begin{align}
\nu=\sqrt{J_\tau^2-J_0^2}
\label{eq2:nubif}
\end{align}
This solution exists provided $J_\tau^2>J_0^2$.
When this condition holds, a Hopf bifurcation arises in which the steady state changes stability while the imaginary part $\nu$ of the eigenvalue is non-zero. Such a bifurcation leads to an oscillatory, limit-cycle behavior with frequency $\nu$ at the bifurcation.

In order to make sense of the condition $J_\tau^2>J_0^2$ given above, we now go to the limit of large $\delta$ (more explicitly, $\delta\gg C/\beta, C/\gamma$).
This assumption is experimentally reasonable, since stress release in the biofilm can be expected to be fast, as it depends on the opening of ion channels \cite{Prindle:2015uq} and on metabolic processes that operate on the order of minutes, whereas the period of the oscillations is on the order of hours.
In this case $x_s \approx 0$. This can intuitively be seen by considering Eq.~\ref{eq:stdst} as the crossing of the function $f(x)$ (see Fig.~\ref{fig2}B), which has a value of $C$ at $x=0$, with the line $y=\delta x$, which has an increasingly higher slope as $\delta$ increases, and in the limit of a vertical line it would intersect the function at $x_s=0$. Since, according to Eq.~\ref{eq:Jtau}, $J_\tau=-\alpha$ for $x_s=0$, we can rewrite the condition for the existence of the Hopf bifurcation as $\alpha>\delta$. Therefore, for strong enough negative feedback the system exhibits a Hopf bifurcation leading to oscillations.

\section*{The Hopf bifurcation is subcritical}

To further characterize analytically the Hopf bifurcation exhibited by the system, we now rewrite Eqs.~\ref{eq:dde}-\ref{eq:fx} as
\begin{align}
\epsilon\,\frac{dx}{dt}= \widetilde{C}-\frac{ \widetilde{\alpha}\,x(t-\tau)}{1-\left(x(t-\tau)/\beta\right)^2+\left(x(t-\tau)/\gamma\right)^4}-x\,,
\label{model_}
\end{align}
where we define $\epsilon=\frac{1}{\delta}, \widetilde{C}=\frac{C}{\delta}$, and $\widetilde{\alpha}=\frac{\alpha}{\delta}$.
Taking again the limit of large $\delta$ ($\epsilon\to 0$), the system reduces to the discrete map $x_{n+1} = \widetilde{f}(x_n)$ \cite{larger2004subcritical}, where
\begin{align}
\widetilde{f}(x)=\widetilde{C}- \frac{\widetilde{\alpha}x}{1-{x}^{2} + x^{4}}
\end{align}
assuming again $\beta=\gamma=1$.
In this discrete description, the condition for a Hopf bifurcation to occur at a critical value $ \widetilde{\alpha}_c$ is \cite{strogatz2014nonlinear}:
\begin{align}
\widetilde{f}_x\left( \widetilde{\alpha}_c,\widetilde{x}_s( \widetilde{\alpha}_c)\right)=-1\,,
\label{eq:Hopfcond}
\end{align}
where the subindex `$x$' in $\widetilde{f_x}$ indicates differentiation with respect to $x$, and $\widetilde{x}_s( \widetilde{\alpha}_c)$ denotes the steady state evaluated at $\widetilde{\alpha}_c$.
Assuming without loss of generality that $\widetilde{C}= 0$, the fixed point of the map such that $x_n=\widetilde{f}(x_n)$ is $\widetilde{x}_s=0$, and $\widetilde{f}_x( \widetilde{\alpha}_c,\widetilde{x}_s)=-\widetilde{\alpha}_c$. This implies, according to Eq.~\ref{eq:Hopfcond}, that the critical value of the rescaled feedback intensity is $\widetilde{\alpha}_c=1$, corresponding to $\alpha_c=\delta$ and in agreement with the analysis made in the preceding section.

The stability of the steady state around the Hopf bifurcation point can be established by the following quantity \cite{Larger:2001kq}:
\begin{align}
a=-\left(\widetilde{f}_{xx}\widetilde{f}_\alpha+2\,\widetilde{f}_{x \alpha}\right)
\end{align}
where the subindex `$\alpha$' denotes differentiation with respect to $\widetilde{\alpha}$, and the partial derivatives are to be evaluated at the fixed point $\widetilde{x}_s$ and at the critical value $\widetilde{\alpha}_c$ of the control parameter.
A linear stability analysis shows \cite{Larger:2001kq} that the steady state a small distance $\Lambda= \widetilde{\alpha}-\widetilde{\alpha}_c$ away from the bifurcation point is stable (unstable) if $a\Lambda<0$ ($a\Lambda>0$).
Furthermore, the limit cycle emerging at the Hopf point corresponds in this discrete description to a period-2 fixed point $x_n=\widetilde{f}(\widetilde{f}(x_n))$, with $x_n=\widetilde{x}_s+\sqrt{-a\Lambda/b}$ \cite{larger2004subcritical} and $b$ equal to:
\begin{align}
b=-\left(\frac{1}{2}\widetilde{f}_{xx}^{\,2}+\frac{\widetilde{f}_{xxx}}{3}\right)
\end{align}
Considering again $\widetilde{C}=0$, we find that $a$ and $b$ are both positive ($a=2$ and $b=2$).
Therefore, the limit cycle that emerges at the Hopf bifurcation exists for $\Lambda<0$ ($\widetilde{\alpha}<\widetilde{\alpha}_c$), and coexists with a stable fixed point (since $a\Lambda<0$). We can thus conclude that the Hopf bifurcation exhibited by this system is \textit{subcritical}.

\section*{Oscillations can coexist with a stable steady state}

\begin{figure}[htb]
\centerline{\includegraphics[width=0.65\textwidth]{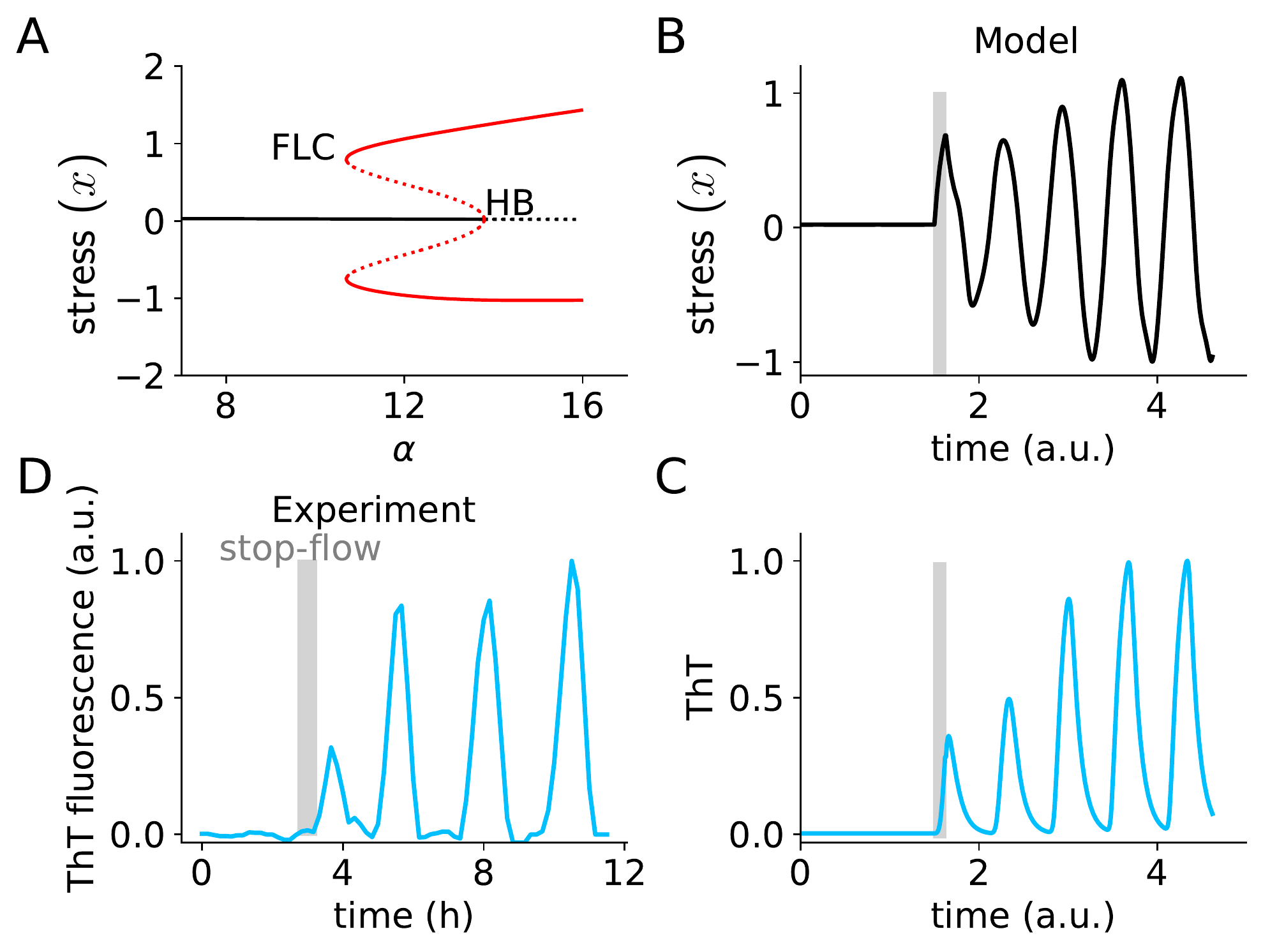}}
\caption{The system has a subcritical Hopf bifurcation. (A) Bifurcation diagram of the model with respect to $\alpha$, for $C=0.5$, $\tau=0.25$, $\beta=\gamma=1$, $\delta=10$. The black lines represent steady states; the red lines denote limit cycles (with solid lines corresponding to stable attractors, and dashed lines to unstable ones).
HB: Hopf bifurcation; FLC: Fold bifurcation of limit cycles.
(B-C) Simulation for $\alpha=12.5$. The grey region denotes a perturbation of parameter $C$, which is increased to 10 for 0.125 time units. The ThT trace is normalised to its maximum. ThT parameters:  $\alpha_T=10, x_{\rm th}=0.6, \sigma=0.1, \delta_T=10$. (D) Experimental ThT time trace showing the response to a 30-min stopping of nutrient flow (grey region). }\label{fig3}
 \end{figure}

In order to determine whether the conclusion above holds for the complete system (with finite $\delta$) we performed continuation analysis using the DDEbiftool software package \cite{ddebiftool1}.
Figure~\ref{fig3}A shows that indeed the system undergoes a subcritical Hopf bifurcation (HB) as $\alpha$ increases.
Moreover, as usual in subcritical bifurcations, at low enough value of $\alpha$ the unstable limit cycle folds back into a stable limit cycle (in what is known as a fold bifurcation of limit cycles, FLC) that coexists with the stable steady state, resulting in a region of \textit{bistability}, where the oscillatory dynamics shown in Fig.~\ref{fig1} coexists with a non-oscillating state in which the biofilm grows in a monotonic manner.
When the system is in the stable steady state within this bistability region, a perturbation (in the form, for instance, of a temporary increase in the basal stress production parameter $C$) can make it jump to the oscillatory attractor.
This can be observed in Fig.~\ref{fig3}B, which shows the response of the stress variable $x$ to a brief perturbation (represented by the vertical gray region in the plot).

To ease comparison with experiments, we model the dynamics of ThT as a reporter located downstream of $x$. To that end, we assume that ThT accumulates in the cell as $x$ increases in a sigmoidal manner with threshold $x_{\rm th}$, and decays linearly \cite{Prindle:2015uq}:
\begin{align}
\frac{dT}{dt}=\frac{\alpha_T}{\exp[(x_{\rm th}-x)/\sigma]+1}-\delta_T\,T
\end{align}
Fig.~\ref{fig3}C shows the dynamics of ThT corresponding to the $x(t)$ time trace in Fig.~\ref{fig3}B.
The plot confirms that ThT responds in a bistable manner, remaining in a (stable) stationary solution until it is perturbed, at which point it jumps to the coexisting oscillatory attractor.

According to the bifurcation behavior described above, we should expect that perturbing experimentally an otherwise stationary biofilm should lead to oscillations.
To validate this expectation experimentally we use the fact that, in our setup, nutrients (in particular glutamate) are flowing through the microfluidic device in a continuous manner.
We can thus perturb the biofilm by stopping the flow temporarily, which leads to a sudden increase in glutamate starvation and stress, and is thus analogous to increasing the $C$ parameter in the model.
We call this a ``stop-flow'' perturbation in what follows.
Figure~\ref{fig3}D shows the response of the biofilm to such perturbation (again represented by a vertical gray bar).
In agreement with the subcritical nature of the bifurcation exhibited by the model, transiently stopping the flow in growing biofilms triggers the emergence of oscillations, which quickly reach their final amplitude.
We thus conclude that the stress oscillations exhibited by the biofilm coexist in a bistable manner with a non-oscillating state, as expected from the delay-differential model proposed above.

\section*{Oscillations require a minimum system size}

\begin{figure}[htb]
\centerline{\includegraphics[width=0.65\textwidth]{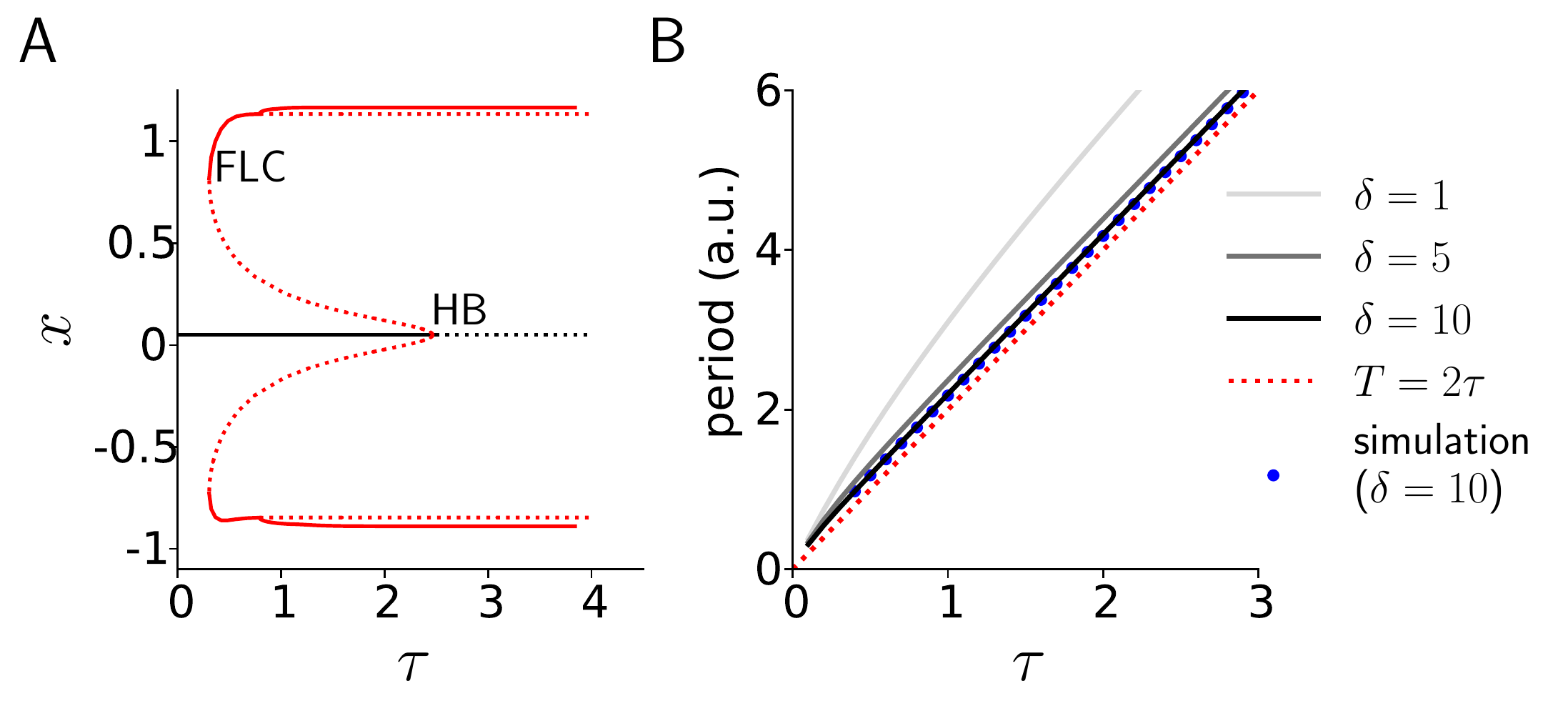}}
 \caption{The subcritical Hopf bifurcation can be controlled by the time delay ($\tau$). (A) Bifurcation diagram as a function of $\tau$, for $C=1, \alpha=10, \beta=1, \gamma=1, \delta=10$.
 (B) Analytic relationship between the oscillation period and the time delay for different values of $\delta$ (solid lines).
 The red dotted line shows the limit of infinite $\delta$, and the blue dots show simulation results for $\delta=10$.
 }\label{fig4}
 \end{figure}

As biofilms grow their size increases, and therefore the parameter that drives the system to the bifurcation point should be the delay $\tau$ rather than $\alpha$.
Solving Eqs.~\ref{eq1:real}-\ref{eq2:im} for $\tau$ with $\mu=0$ leads to the exact critical value of the delay at which the bifurcation occurs:
\begin{align}
\tau_c=\sqrt{\frac{\arccos(-J_0/J_\tau)^2}{J_{\tau}^2-J_0^2 }}=\sqrt{\frac{\arccos(\delta/J_\tau)^2}{J_{\tau}^2-\delta^2 }}\,,
\label{tcrit_d}
\end{align}
which, again, exists provided $J_\tau^2 > \delta^2$.
Continuation analysis confirms that the system can also undergo a subcritical Hopf bifurcation when increasing $\tau$, as shown in Fig.~\ref{fig4}A. The figure shows that there is also a large region of bistability in this case. Thus, the model predicts that for small delays the system is at a steady state, it then enters a bistable region and finally crosses the bifurcation point, where only an oscillatory steady state remains. Incidentally, period doubling bifurcations arise when $\tau$ increases (see branching of red line in Fig.~\ref{fig4}A; higher-order doublings arise for larger values of $C$, results not shown).

We can also explore theoretically the relationship between the period of the oscillations and the time delay.
To that end, we assume that the oscillation frequency will be given approximately by the imaginary part $\nu$ of the eigenvalue of the steady state.
Combining Eqs.~\ref{eq1:real} and \ref{eq2:im} through the elimination of $\exp(-\mu\tau)$ leads to the following equation for $\nu$:
\begin{align}
0=\frac{-\nu}{\delta+\mu}-\tan(\nu\tau) \approx \frac{-\nu}{\delta}-\tan(\nu\tau),
\label{nu_mu}
\end{align}
where we assume again the limit of large $\delta \gg \mu$. In that limit, the equation above has three solutions for $\nu\tau$ in $[-\pi,\pi]$, which are asymptotically close to $0$, $\pi$ and $-\pi$.
Taking into account that $J_\tau\approx -\alpha<0$ for large $\delta$, and that $J_0=-\delta<0$ in any case, Eqs.~\ref{eq1:real} and \ref{eq2:im} lead to the conclusion that $\cos(\nu\tau)<0$ (and close to $-1$) and $\sin(\nu\tau)>0$ (and close to $0$).
The only one of the three solutions listed above that fulfils these conditions is $\nu\tau\approx\pi$. Therefore, in the limit of large $\delta$ we can expect the oscillation period to be $T\sim2\pi/\nu\approx 2\tau$.
This expression is shown as a red dotted line in Fig.~\ref{fig4}B.
The figure also plots the result of numerically solving Eq.~\ref{nu_mu} for $\alpha=10$ and different finite values of $\delta$.
For $\delta=10$, for instance, the relationship between the period and the delay lies slightly above the line $T= 2\tau$, and matches the numerical simulations of the system for those parameter values.
This linear relationship between the period and the delay is consistent with the experimental observations (Fig.~\ref{fig1}D), further validating our interpretation of the time delay as being directly related with the biofilm size.

\begin{figure}[ht!]
\centerline{\includegraphics[width=0.65\textwidth]{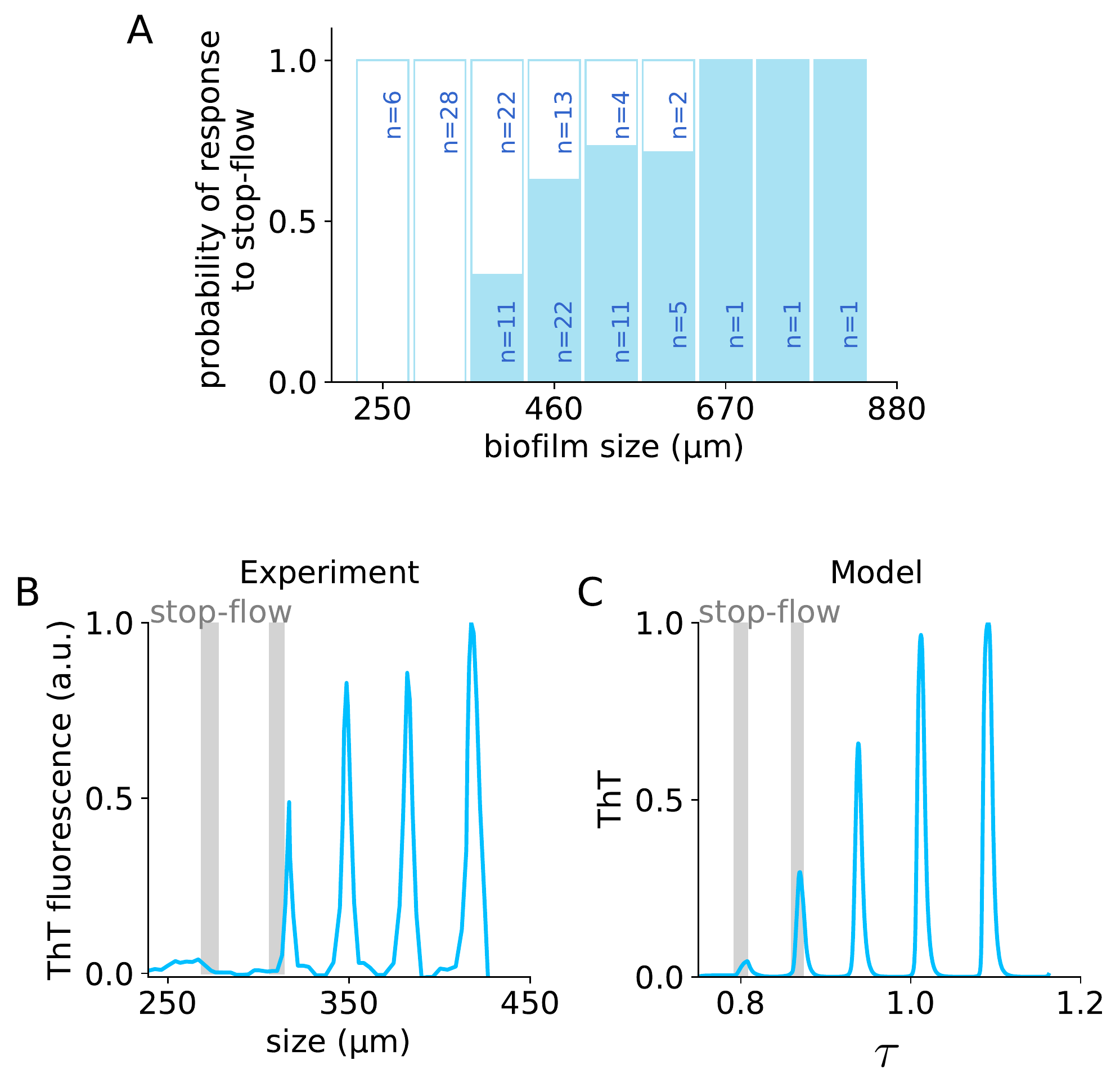}}
\caption{Oscillation triggering in the bistable regime is easier for larger biofilms.
(A) Fraction of experiments (solid blue) in which a stop-flow perturbation triggers oscillations in growing biofilms.
The flow was stopped for 30 min and the appearance of oscillations was assessed. Non-responding biofilms are indicated in the white bars.
(B) Sample ThT experimental time trace where an early application of a stop-flow perturbation (vertical grey bar) does not trigger oscillations, but a subsequent one when the system is slightly larger does.
(C) Simulation of the system with increasing $\tau$. Two subsequent stop-flow-type perturbations (increase of $C$ to 3.5) were applied, at times 1.25 and 3.25, for 0.5 time units. Parameter values are those of Fig.~\ref{fig4}, plust $\eta=0.035$, $\delta_{\tau}=0.02$, $\alpha_T=10$, $x_{\rm th}=0.6$, $\sigma=0.1$, $\delta_T=10$. Initial conditions are $\tau_0=0.75$, $T_0=0$, $x_0=0.05$. ThT time trace is normalised to its maximum. }
\label{fig5}
\end{figure}

The bifurcation behavior shown in Fig.~\ref{fig4}A indicates that beyond a critical delay (corresponding to a critical biofilm size), the only stable attractor of the system is an oscillatory one.
However, even before that critical size is reached, the biofilm can be induced to oscillate if properly perturbed (as shown in Fig.~\ref{fig3} above).
The `ease' with which a perturbation will induce the biofilm to jump from steady to oscillatory growth within the bistable region is (roughly) inversely related with the size of the basin of attraction of the stable fixed point.
An indication of the size of this basin of attraction is provided by the amplitude of the unstable limit cycle (dashed red line in Fig.~\ref{fig4}A) surrounding the stable fixed point (solid black line in the plot).
The bifurcation diagram shows that the unstable limit cycle gets closer to the stable fixed point as $\tau$ (and hence the system size) increases.
Therefore, the model predicts that the closest the system is to the bifurcation point, the easiest it will be that a perturbation drives it out of the stable fixed point and into the oscillatory attractor.
This agrees well with the experimental data shown in Fig.~\ref{fig5}A, where larger biofilms are more likely to start oscillating upon a stop-flow perturbation.
Furthermore, biofilms under a minimum size of around 400~$\mu$m never oscillate, and those beyond a maximum size of around 670~$\mu$m always oscillate.

Since the biofilm is continuously growing, the responsiveness to perturbations depends on time: perturbing the biofilm when it is too small does not result in oscillations, whereas a subsequent perturbation does (Fig.~\ref{fig5}B).
In order to model this behavior and explore the interplay between growth and consecutive perturbations, we assume that the time delay in Eq.~\ref{eq:dde} increases at a rate that diminishes with $x$, since when the biofilm is stressed above (below) the basal level, its growth rate will decrease (increase), and so will its size, and correspondingly the delay:
\begin{align}
\frac{d\tau}{dt}=\eta-\delta_{\tau}x
\end{align}
(Parameters $\eta$ and $\delta_{\tau}$ are chosen in such a way that $d\tau/dt$ is always positive).
We used this dynamics to investigate the response to two consecutive perturbations in the model.
According to the simulations, if the system is too small, a stop-flow perturbation may not perturb the system sufficiently to make it jump to the oscillatory attractor (Fig.~\ref{fig5}C).
But a subsequent perturbation when the delay is larger can, as in the experiments (compare Figs.~\ref{fig5}B,C).

\section*{The critical size for oscillation onset is a function of the stress level}

\begin{figure}[htb]
\centerline{\includegraphics[width=0.65\textwidth]{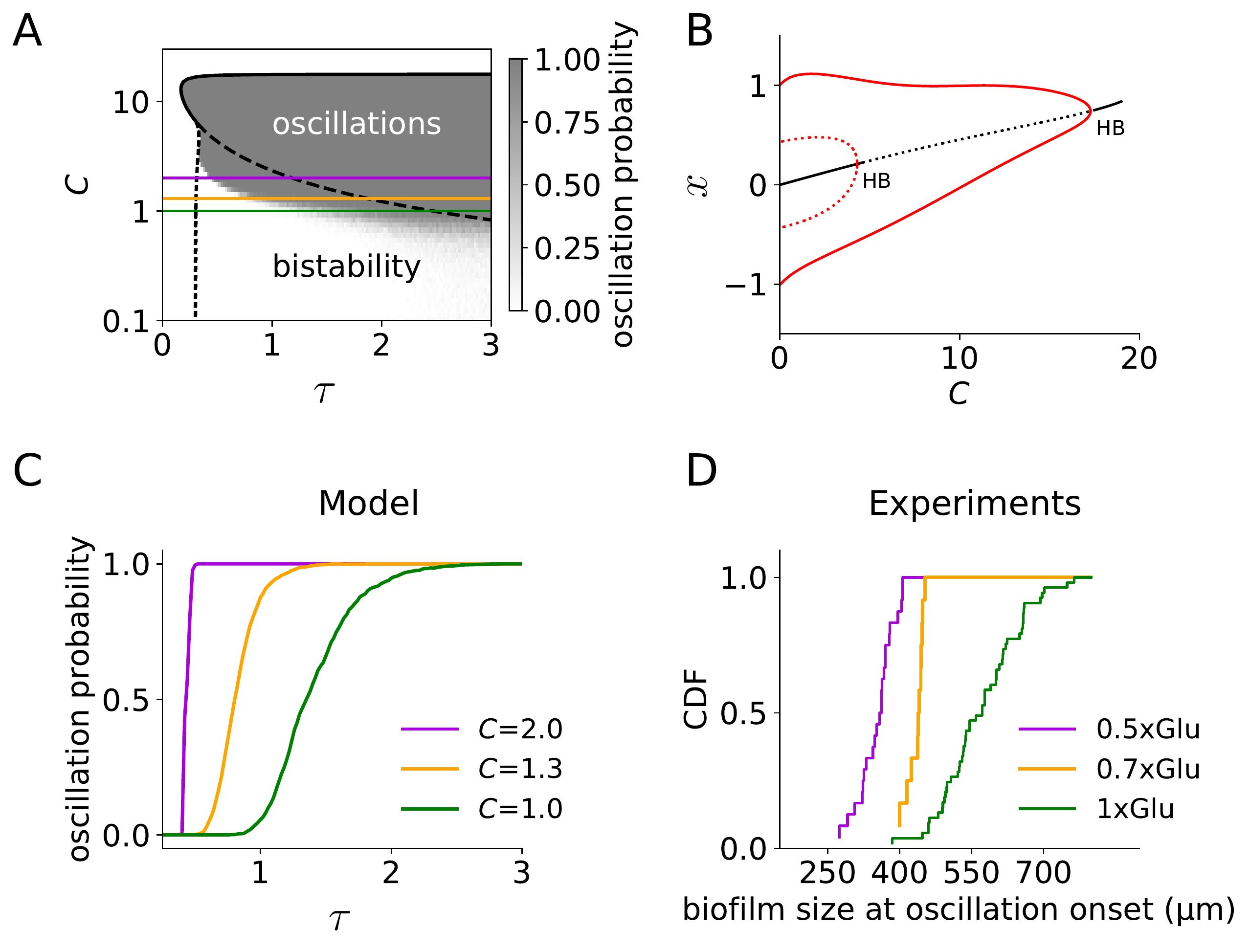}}
\caption{Stress reduces the critical size for oscillations. (A) Phase diagram showing the position of the Hopf (solid and dashed black lines) and fold bifurcation of limit cycles (dotted black line), as a function of the basal stress production $C$ and the feedback delay $\tau$. The colormap shows the probability of oscillation in stochastic simulations with stop-flow consisting on an increase in $C$ by a factor of 3 during 0.5 time units. Parameters are those of Fig.~\ref{fig4}.
(B) Bifurcation diagram as a function of $C$ for fixed $\tau$, corresponding to a vertical cross-section of panel A at $\tau= 0.5$.
(C) Horizontal cross-sections of the grey field from panel A at different $C$ levels (smoothed).
(D) Experimental  cumulative distribution functions (CDF) of the size at oscillation onset for three different glutamate concentrations.}
\label{fig6}
\end{figure}

As shown in Fig.~\ref{fig4}A, the bistability region is bounded on the right by the critical delay at which the Hopf bifurcation (HB) occurs, and on the left by the point at which the unstable limit cycle becomes stable at the fold bifurcation of limit cycles (FLC).
We performed a continuation analysis to explore how these two bifurcation points change as a function of the delay $\tau$ and the basal stress production $C$.
Figure~\ref{fig6}A represents the resulting phase diagram, where the purely oscillatory region is bound by the Hopf bifurcation line (thick black line, with solid representing supercriticality and dashed subcriticality), and the bistable region is delimited by the subcritical HB line and the FLC line (dotted black line).
A 1-d vertical cross section of this phase diagram for fixed $\tau$ is shown in Fig.~\ref{fig6}B as a bifurcation diagram with respect to $C$, and confirms that the entire region below the subcritical HB is indeed bistable.
The phase diagram depicted in Fig.~\ref{fig6}A shows that the bistability region is wide.
Moreover, the analysis indicates that for high enough values of $C$ the Hopf bifurcation as a function of $\tau$ becomes supercritical and bistability is lost.
At this point, $f(x)$ is highly biased towards positive values, mostly contributing to stress increase.
This suggests that bistability requires the system to relieve stress sufficiently below baseline levels.

Within the bistable region, Fig.~\ref{fig6}A also predicts that for higher basal stress levels (higher $C$), the critical delay at which the Hopf bifurcation happens (dashed line) is reduced, while the position of the fold bifurcation remains unchanged (dotted line).
We confirm this result by adding noise to the system and integrating numerically the corresponding stochastic delay-differential equation (see Materials and Methods) for different $C$ and delay values.
We compute the response of the system to a temporary increase of the basal stress production rate $C$ for 100 realizations of the noise, and quantify the fraction of simulations in which the system switches from the stable fixed point to the limit cycle attractor.
In agreement with the continuation analysis, we found that for higher basal stress levels the system is able to respond to the perturbation for smaller delays (Fig.~\ref{fig6}C).
We can validate this prediction experimentally by tuning the glutamate concentration of the medium, with lower concentrations being associated with higher values of $C$.
Identifying the cumulative distribution of sizes at which biofilms are found to oscillate (either naturally or upon a stop-flow perturbation) with the propensity of the biofilm to switch to oscillations for a given size, we observe (Fig.~\ref{fig6}D) a good qualitative agreement with the model: for higher stress levels (lower glutamate) the critical size to respond to perturbations is reduced.
The model predicts that the range of the bistable sizes is reduced as stress increases, which is also in agreement with the experimental data (compare e.g. the ranges of the magenta and green lines in Figs.~\ref{fig6}C and D).

\section*{Discussion}

We have studied the transition to growth oscillations in bacterial biofilms.
Our experiments show that the oscillations arise for a minimum biofilm size, with a period that increases with that size.
We have proposed a minimal model to explain these features, assuming a general functional behavior of stress production.
The resulting delay-differential equation model shows emergence of oscillations at a critical delay (which we link with biofilm size), through a subcritical Hopf bifurcation.
Such bifurcation entails the presence of a bistable regime in which the biofilms either oscillate in their growth or expand steadily, depending on the initial conditions.
Experimentally, this expectation can be validated by temporarily stopping the media flow within the microfluidic device.
Our experiments do show that such perturbations lead to oscillations in growing biofilms, provided they are large enough.

Biofilm oscillations were described to be a mechanism to prevent cells in the biofilm interior from dying due to starvation caused by the growth of peripheral cells, and this was shown to enable the regeneration of the community upon external chemical attacks \cite{Liu:2015fk}.
Therefore, the bistable behavior reported here could be a mechanism to allow oscillations to start early enough to ensure the survival of interior cells.
Both the model and the data show that the critical size for biofilm oscillations depends on the nutrient concentration in the media.
This is consistent with the fact that when nutrient levels are low, interior cells become starved earlier, thus triggering the emergence of the oscillations at smaller sizes. Also, the discontinuous transition allows the biofilm to start oscillating immediately with a non-zero amplitude, instead of having to wait for the oscillations to slowly develop.

It is interesting to note that our delay-differential model is very similar to the classical Mackey-Glass equation originally proposed to model the production of mature blood cells \cite{Mackey:1977ys}.
In contrast with our case, however, the Mackey-Glass model is known to have a supercritical bifurcation.
The authors also noted that it has period doubling bifurcations and eventually an aperiodic or chaotic regime.
We also observed such richness in dynamical behaviour in our case.
However, although mathematically interesting, we have not further analysed it as it does not seem relevant to our experimental situation.

Delayed feedback has been recognized as a source of dynamical behavior in many areas of science and engineering for years \cite{erneux2009applied}.
Spatially structured biological populations provide a natural substrate for delayed interactions due to the finite propagation speed of biological signals.
The results reported here show that such delayed interactions can have important collective effects in bacterial populations.
It would be interesting to study the roles played by similar delay-induced mechanisms in other biological systems requiring global spatiotemporal coordination, such as developing organisms.

\section*{Materials and Methods}

\subsection{Biofilm culture conditions and stop-flow procedure}
{\em Bacillus subtilis} biofilms were grown in microfluidics chips as described previously \cite{Liu:2015fk,Prindle:2015uq}. Media flow was driven by a pneumatic pump from the CellASIC ONIX Microfluidic Platform (EMD Millipore). The pump pressure was kept stable during the course of the experiments. We used a pump pressure of 1.5 psi with only one media inlet open, which maintains a media flow of ~24 $\mu$m/s in the microfluidic chamber. During each stop-flow perturbation, the pump was turned off for 30 min. Biofilms were monitored using time-lapse microscopy, and we tracked metabolic oscillations within growing biofilms using 10 $\mu$M Thioflavin T (ThT). Experimental time traces are detrended using spline interpolation on the troughs of oscillations, smoothed and normalised to the maximum.

\subsection{Continuation analysis}
Bifurcation diagrams were computed using the DDE-BIFTOOL v.3.1.1 package \cite{ddebiftool1}, run in Octave \cite{octavecit}.

\subsection{Simulations}
Deterministic simulations of the system for fixed values of $\tau$ were carried out using the Python package {\tt pydelay v.0.1.1} \cite{pydelay}.
Deterministic simulations with state-dependent delays were performed using the routine ddesd \cite{ddesd} in Matlab (The MathWorks Inc.). Initial conditions were the analytical steady state for the $x$ variable and 0 for ThT.

In order to introduce noise into the system, we included an additive Gaussian white noise $\xi(t)$ in the stress equation:
\begin{align}
\frac{dx}{dt}=C-\frac{\alpha\,x(t-\tau)}
{1-\left(x(t-\tau)/\beta\right)^2+\left(x(t-\tau)/\gamma\right)^4}
-\delta\,x+D\xi(t),  \label{model_s}
\end{align}
with noise strength $D=0.03$. We integrated this stochastic DDE with a custom-made code by adapting the stochastic Heun algorithm \cite{toral2014stochastic} to include the delay.

To calculate the oscillation probability at each combination of $\tau$ and $C$ values (Fig.~\ref{fig6}A), we performed 100 simulations per parameter combination.
For each simulation, a random initial history array was generated from a uniform distribution in the interval $[0.5x_s,1.5x_s]$, where $x_s$ is the steady state value.
We then integrated Eq.~\ref{model_s} for 100 time units, at which point the $C$ value was increased a factor of 3 for 0.5 time units in order to simulate the stop-flow, and then returned to basal level for 1000 time units.
The last 100 time units were used to classify each trace as oscillatory or not, depending on whether it exhibited oscillations of amplitude larger than one.

\section*{Acknowledgements}
This work was supported by the Spanish Ministry of Economy and Competitiveness and FEDER (project FIS2015-66503-C3-1-P), and by the Generalitat de Catalunya (project 2017 SGR 1054).
R.M.C. acknowledges financial support from La Caixa foundation.
J.G.O. acknowledges support from the ICREA Academia programme and from the ``Mar\'ia de Maeztu'' Programme for Units of Excellence in R\&D (Spanish Ministry of Economy and Competitiveness, MDM-2014-0370). G.M.S. acknowledges support for this research from the San Diego Center for Systems Biology (NIH grant P50 GM085764), the National Institute of General Medical Sciences (grant R01 GM121888 to G.M.S.), the Defense Advanced Research Projects Agency (grant HR0011-16-2-0035 to G.M.S.), the Howard Hughes Medical Institute-Simons Foundation Faculty Scholars program (to G.M.S.).

\bibliography{subdelay}

\end{document}